\documentclass[twocolumn]{aastex61}
\received{ }
\revised{}
\accepted{}
\submitjournal{ApJ}

\shorttitle{ A new oxygen-rich Mira variable}
\shortauthors{Ghosh et al.}
\usepackage{ulem}

\begin{document}

\title{Phase-dependent photometric and spectroscopic characterization of the MASTER-Net Optical Transient J212444.87+321738.3: an oxygen rich Mira}

\correspondingauthor{Soumen Mondal}
\email{ soumen.mondal@bose.res.in, supriyo12a@boson.bose.res.in}

\author[0000-0002-0786-7307]{Supriyo Ghosh}
\affil{S. N. Bose National Centre for Basic Sciences, Salt Lake, Kolkata-700 106, India}

\author{Soumen Mondal}
\affiliation{S. N. Bose National Centre for Basic Sciences, Salt Lake, Kolkata-700 106, India}
%\nocollaboration{(AAS Journals Data Scientists collaboration)}

\author{Ramkrishna Das}
\affiliation{S. N. Bose National Centre for Basic Sciences, Salt Lake, Kolkata-700 106, India}
%\affiliation{AAS Journals Associate Editor-in-Chief}
%\nocollaboration

\author{D. P. K. Banerjee}
\affiliation{Physical Research Laboratory, Navrangpura, Ahmedabad-380 009, India}
%\collaboration{(LaTeX collaboration)}

\author{N.M. Ashok}
\affiliation{Physical Research Laboratory, Navrangpura, Ahmedabad-380 009, India}

\author{Franz-Josef Hambsch}
\affiliation{Vereniging Voor Sterrenkunde (VVS), Brugge, BE-8000, Belgium}
\affiliation{American Association of Variable Star Observers (AAVSO), Cambridge, USA}

\author{Somnath Dutta}
\affiliation{S. N. Bose National Centre for Basic Sciences, Salt Lake, Kolkata-700 106, India}



\begin{abstract}

We describe the time-dependent properties of a new spectroscopically confirmed Mira variable, which was discovered in 2013 as MASTER-Net Optical Transient (OT) J212444.87+321738.3 towards the Cygnus constellation. We have performed long-term optical/near-infrared (NIR) photometric and spectroscopic observations to characterize the object. From the optical/NIR light curves, we estimate a variability period of 465 $\pm$ 30 days. The wavelength-dependent amplitudes of the observed light-curves range from $\Delta$I$\sim$4 mag to $\Delta$K$\sim$1.5 mag. The (J-K) color-index varies from 1.78 to 2.62 mag over phases. Interestingly, a phase lag of $\sim$60 days between optical and NIR light curves is also seen, as in other Miras. Our optical/NIR spectra show molecular features of TiO, VO, CO, and strong water bands which are a  typical signature of oxygen-rich Mira. We rule out S- or C-type as ZrO bands at 1.03 and 1.06 $\mu$m and $C_2$ band at 1.77 $\mu$m are absent. We estimate the effective temperature of the object from the SED, and distance and luminosity from standard Period-Luminosity relations. The optical/NIR spectra display time-dependent atomic and molecular features (e.g. TiO, NaI, CaI, H$_2$O,CO), as commonly observed in Miras.  Such spectroscopic observations are useful for studying pulsation variability in Miras.    

\end{abstract}


\section{Introduction}

Mira-type variables are in the Asymptotic giant branch (AGB) phase, which is the last stage of stellar evolution before turning into planetary nebulae. Miras are long-period (100$-$1000 days) pulsating variable with a large visible amplitude of more than 2.5 mag. These giants have initial masses $\approx$ 0.8$-$8 $M_\odot$ (low to intermediate main sequence mass), and are generally surrounded by circumstellar matter from huge mass loss rates of $\sim$ $10^{-8}$ $-$ $10^{-4}$ $M_\odot$ $yr^{-1}$ \citep{Jura1990, Habing1996, Mattei1997, Olofsson2004, Herwig2005}. Mira variables have a low effective temperature ($<$ 3500 K), cool extended atmospheres (radius up to few 100$R_\odot$), and luminosity can reach up to a few $10^3L_\odot$ \citep{Mattei1997}. High luminous Mira variables play a significant role in the studies of stellar evolution, stellar populations, galactic- extragalactic structure and evolution \citep{Lancon1999, Dejonghe1999, Groenewegen2009}, and enriches the interstellar medium (ISM) significantly through the high mass-loss \citep{Habing1996}. The high mass loss and relatively low surface temperature of these evolved stars provide a habitable zone for several molecules such as TiO, VO, $H_2$O, and CO in their extended atmospheres. These molecules play important roles in the spectral appearance of Mira variable stars at visual and NIR wavelengths \citep{Lancon2000, Gautschy2004, Aringer2009, Nowotny2010}.

The stars in the AGB phase are radially pulsating and become unstable. Mira variables are thought to be fundamental mode pulsators (e.g., \citealt{ Wood1999, Ita2004}). The pulsation mode of Miras is a function of period, mass and radius \citep{Wood1996}. The Mira pulsation is thought to be originating from the variable ionization zone of hydrogen and helium below the stellar surface (e.g., \citealt{Keeley1970}). From each pulsation cycle, shock waves are generated under the photospheric surface, which in turn create a very complex and dynamic atmosphere \citep{Reid2002}. The photometric light curves of Mira variables represent the oscillating behavior of brightness, surface temperature, radius, atmospheric structure and opacity as the star pulsates \citep{LeBertre1992, Castelaz2000}. The significant visual variation is attributed to the opacity variation of metal oxides in the Mira atmosphere \citep{Reid2002}. The observed radii of Mira stars significantly differ at different optical/NIR wavelengths as seen from high angular resolution observations (e.g., \citealt{Thompson2002}; \citealt{Ireland2004}; \citealt{Perrin2004}; \citealt{Mondal2005}). These results indicate the presence of molecular layers above the continuum-forming photosphere \citep{Wittkowski2008}. The theoretical hydrodynamic pulsation models have been developed to understand the pulsation and mass loss mechanisms, which can predict the time-dependent structure and temporal variations over multiple cycles \citep{Fleischer1992, Bessell1996, Winters1997, Hofner1998, Loidl1999, Winters2000, Tej2003a}. These models are characterized by pulsation-driven shocks, non-equilibrium chemistry, and formation of dust grains \citep{Bieging2002}. 

Spectroscopic observational studies are valuable tools to understand pulsating atmospheres and mass loss of Mira variables throughout the outer layer \citep{Hinkle1982, Alvarez2000, Castelaz1997, Castelaz2000, Lancon2000, Tej2003b}. Phase-dependent spectroscopic studies are very efficient to probe the atmosphere of the stars, and such studies are limited in the literature. From radial velocity, excitation temperatures and lines broadening measurements using high resolution spectroscopy, it is evident that two separate line forming regions of the atmosphere sometimes contribute towards the spectrum of H$_2$O (except maximum light), CO ($\Delta \nu=2$), and OH \citep{Hinkle1978, Hinkle1979a, Nowotny2010}. More complex stratification may also exist (e.g., \citealt{Tej2003a, Tsuji2009}).

In this paper, we have studied an object, which was first detected from  MASTER Optical Transient (OT) alert on J212444.87+321738.3 (hereafter, J2124+32) on 2013 March 13 with 10.7 mag at unfiltered CCD \citep{Tiurina2013}.
The Mobile Astronomical System of TElescope Robots (MASTER) Global Robotic Net\footnote{\url{http://observ.pereplet.ru/}} consists of several identical observing instruments at different observatories (e.g., MASTER-Amur, MASTER-Tunka, MASTER-SAAO). The facilities provide very fast sky-survey (128 $deg^2$ per hour) with limiting magnitude 19--20 \citep{Lipunov2016, Lipunov2017}. The primary goals of the MASTER-net are to observe gamma-ray bursts (GRB) in alert mode. However, it discovers many OTs in the survey mode. The other identification of object are USNO-B1.0 1222-0647260, 2MASS 21244500+3217377 and WISE J212444.98+321737.73.440 \citep{Tiurina2013}. Following the OT announcement, we have started spectro-photometric monitoring observations on the object in optical/NIR wavelength since 2013 March 20 for more than 1.5 years using different telescope facilities. We present here the result of characterization MASTER OT J2124+32, which turns out to be a new O-rich Mira variable. From the 3rd and 4th edition of General Catalog Variable Stars, galactic Mira detection limit is complete down to maximum magnitude V $\approx$ 9  mag. \citep{Kharchenko2002}. This new Mira variable has a peak magnitude at I-band $\approx$10.4 mag, which corresponds to V$\approx$14 \citep{Kharchenko2002}. Because of faintness,  the object might not be included in the variability monitoring program. The paper is organized as we describe the details of our observations and data reduction procedures in section II and section III deals with our new results and discussion. Finally, the summary and conclusion of our studies are mentioned in section IV.

\begin{table*}
\begin{center}
\caption{ Log of photometric and spectroscopic observations}
\label{tab:log}
\resizebox{1.0\textwidth}{!}{
\begin{tabular}{crrrrrrr}
\hline
Date of Observation & Observation Type & Spectral Band & Int. Time (s) & No of Frames & Telescope & Remarks\\
\hline
\hline
 2013-Mar-20& Photometry& J/H/K& 0.4/0.2/0.2 &5* [21/21/21]  & 1.2m Mt. Abu& clear sky\\ 
 2013-Mar-22& Photometry& J/H/K&0.5/0.4/0.2 &5* [11/11/21]  & 1.2m Mt. Abu & clear sky\\   
 2013-Apr-28& Photometry & J/H/K & 2/1/0.3&5* [15/15/25] &1.2m Mt. Abu & clear sky \\ 
 2013-Apr-29& Photometry & J/H/K & 0.5/0.5/0.2&5* [31/31/31] &1.2m Mt. Abu & clear sky \\ 
 2013-May-27& Photometry & J/H/K &0.3/0.7/1 &5* [15/15/35] &1.2m Mt. Abu & clear sky  \\ 
 2013-May-28& Photometry & J/H/K & 1/1/0.3&5* [21/21/35] &1.2m Mt. Abu & clear sky \\ 
 2013-May-30& Photometry & J/H/K &0.3/0.5/1 &5* [15/15/15] &1.2m Mt. Abu & clear sky\\ 
 2013-Jun-20& Photometry & J/H/K &0.3/0.5/1 &5* [15/15/15] &1.2m Mt. Abu & clear sky\\   
 2013-Oct-30& Photometry & J/H/K &0.3/0.5/1 &5* [15/15/15] &1.2m Mt. Abu & clear sky\\   
 2015-May-08& Photometry & J/H/K &0.3/0.5/1 &5* [15/15/15] &1.2m Mt. Abu & clear sky\\ 
 \hline
 2013-Apr-02 & Photometry & I/CV & 60/30 & 515/515 & 40cm Chile & ---\\ 
 - 2014-Aug-31 & & & & & & \\
 \hline
 2013-Mar-20& Spectroscopy & J/H/K/KA & 120/90/60/60 & 2*1 &1.2m Mt. Abu & clear sky \\
 2013-Apr-28 & Spectroscopy & J/H/K/KA & 90/75/60/60 & 2*1 & 1.2m Mt. Abu 
& clear sky \\
2013-May-30 & Spectroscopy  & J/H/K/KA & 90/60/60/60 & 2*1 & 1.2m Mt. Abu
 & clear sky \\
 2013-Oct-15& Spectroscopy & 600-920~$nm$ &1800 &1 & 2m HCT & clear sky\\ 
2014-May-21 & Spectroscopy & 600-920~$nm$ &300 & 1 & 2m HCT & clear sky\\
2014-Aug-18 & Spectroscopy &   H/K &30/30/20/10  & 2*5 & 2m HCT & clear sky \\ 
2014-Aug-19 & Spectroscopy &600-920~$nm$ &300 & 1 & 2m HCT & clear sky \\
2014-Oct-06 & Spectroscopy &YJ/HK &100/100 & 2*7  & 2m HCT & clear sky \\
            &              &600-920~$nm$ &600 & 1 &        &        \\
2014-Dec-12 & Spectroscopy &YJ/HK &100/100 & 2*7  & 2m HCT & clear sky\\
            & 	           &600-920~$nm$&600 &1   &    & \\               
2015-Jan-13 & Spectroscopy &YJ/HK &100/100& 2*7   & 2m HCT & clear sky \\
2015-Jan-18 & Spectroscopy &600-920~$nm$&900&1    &2m HCT & clear sky\\
2015-July-05& Spectroscopy &YJ/HK &100/100& 2*5   &2m HCT & clear sky\\
            &              &600-920~$nm$  &1      & 1      & \\
2015-Aug-11 & Spectroscopy &YJ/HK &100/100 &2*5   &2m HCT & clear sky \\
2016-Dec-19 & Spectroscopy &HK    &100     &2*5   &2m HCT & clear sky \\
\hline
\end{tabular}}
\end{center}
\end{table*}

\begin{table*}
\begin{center}
\caption{Near-IR $JHK^\prime$ Photometry}
\label{tab:phot}
\resizebox{1.0\textwidth}{!}{
\begin{tabular}{ccccccc}
\hline
Date of Obs. & Optical Phase & Telescope/Instrument & J & H & K & (J-K)   \\
(UT) & & & (mag) & (mag) & (mag) & (mag) \\
\hline
\hline \\
2013 Mar 22.02 & 0.049  & Mt. Abu/NICMOS-3 & 5.852 $\pm$ 0.055 & 4.467 $\pm$ 0.062  & 3.765 $\pm$ 0.049 & 2.087 \\
2013 Mar 24.01 & 0.051  & Mt. Abu/NICMOS-3 & 5.642 $\pm$ 0.044 & 4.483 $\pm$ 0.035  & 3.765 $\pm$ 0.034 & 1.877 \\
2013 Apr 28.93 & 0.118  & Mt. Abu/NICMOS-3 & 5.668 $\pm$ 0.043 & 4.654 $\pm$ 0.052  & 3.751 $\pm$ 0.010 & 1.917 \\
2013 Apr 29.98 & 0.119  & Mt. Abu/NICMOS-3 & 5.557 $\pm$ 0.021 & 4.510 $\pm$ 0.023  & 3.720 $\pm$ 0.012 & 1.837 \\
2013 May 27.92 & 0.169  & Mt. Abu/NICMOS-3 & 5.621 $\pm$ 0.051 & 4.466 $\pm$ 0.063  & 3.827 $\pm$ 0.045 & 1.794 \\
2013 May 28.95 & 0.170  & Mt. Abu/NICMOS-3 & 5.718 $\pm$ 0.047 & 4.452 $\pm$ 0.053  & 3.752 $\pm$ 0.048 & 1.966 \\
2013 Jun 20.83 & 0.213  & Mt. Abu/NICMOS-3 & 5.807 $\pm$ 0.034 & 4.621 $\pm$ 0.024  & 4.021 $\pm$ 0.029 & 1.786 \\
2013 Oct 30.50 & 0.450  & Mt. Abu/NICMOS-3 & 7.110 $\pm$ 0.020 & 5.660 $\pm$ 0.030  & 4.850 $\pm$ 0.020 & 2.260 \\
2015 May 08.91 & 1.60  & Mt. Abu/NICMOS-3 & 7.868 $\pm$ 0.017 & 6.426 $\pm$ 0.06   & 5.243 $\pm$ 0.015 & 2.625 \\
1998 Jun 18.00\footnote{\citet{Cutri2003}} & & 2MASS/NICMOS    & 7.825 $\pm$ 0.026 & 5.971 $\pm$ 0.017  & 4.818 $\pm$ 0.015 & 3.007 \\
\hline 
Average & & Mt. Abu/NICMOS3 & 6.093$\pm$0.257 & 4.859$\pm$0.220 & 4.077$\pm$0.17 & 2.016 \\
Amplitude (light curve) & & Mt. Abu/NICMOS3 & 2.2 & 1.9 & 1.5 &  \\
\hline
\end{tabular}}
\end{center}
\end{table*}

\section{Observations and Data Reduction}

Optical imaging observations of J2124+32 were performed using a front-illuminated 4K $\times$ 4K CCD camera on the 40 cm f/6.8 Optimized Dall Kirkham (ODK) telescope at the private observatory ROAD (Remote Observatory Atacama Desert) in Chile \citep{Hambsch2012}. The source was monitored in optical I-band and unfiltered CCD (400--900 $nm$) filter-band (Clear, C) over 550 days during 2013 April 02 to 2014 August 31. The accuracy of these observations varies from 0.004 at I=10.40 to 0.047 at I=14.30, while from 0.007 at C= 12.35 to 0.045 at C=15.70 mag respectively. 

The NIR photometric and spectroscopic observations were carried out using Near-Infrared Imaging Camera cum Multi-Object Spectrograph (NICMOS-3) on 1.2m Mt. Abu telescope, India, and TIFR Near-Infrared Spectrometer and Imager (TIRSPEC) on 2m Himalayan Chandra Telescope (HCT) at Hanle, India. The NICMOS3 has 256 $\times$ 256 HgCdTe detector array and provides a resolution R $\approx$1000; while TIRSPEC has 1024 $\times$1024 Hawaii-1 array and provides resolution R $\approx$1200. The spectral coverage of NICMOS-3 were on  $JHK$-bands. The spectrograph in the  NICMOS-3  instrument does not  cover the whole K-band in a single shot. We observed in two parts, the $1^{st}$ part covers 1.9--2.3  $\mu$m ( termed as K), and the $2^{nd}$ part covers 2.1--2.4 $\mu$m with some overlap  (termed as KA). The TIRSPEC spectra were taken at $YJHK$ bands. More details of TIRSPEC could be found elsewhere \citep{Ninan2014}. Photometric observations in $JHK^\prime$-bands were taken in five dithered positions, and multiple frames are taken in each dithered position to get better signal to noise ratio (SNR). In spectroscopic observing mode, the spectra were taken at two different positions along the slit one after another immediately to subtract the sky, and several frames were observed to improve SNR. We have estimated the SNR of our spectra. The SNR  is $\sim$50 (J-band),$\sim$80 (H-band), $\sim$80-100 (K-band) for TIRSPEC data. While the SNR is $\sim$30 (JHK)for NICMOS-3 data. A log of our observations is mentioned in Table~\ref{tab:log}.

The optical spectra are taken using Himalaya Faint Object Spectrograph and Camera (HFOSC) on the 2m HCT at Hanle, India. The HFOSC instrument has several grisms covering different wavelength range and resolution, and we have used Grism no. 8 (Gr\#8) for our studies, which covers the wavelength range of 580-920~$nm$ respectively and provides a resolution R $\approx$ 2200\footnote{\url{https://www.iiap.res.in/iao_hfosc}}.

The data reduction was performed with the help of standard tasks of the Image Reduction and Analysis Facility (IRAF\footnote{\url{http://iraf.noao.edu/}}). In NIR photometric reduction, the sky frames were generated with all dithered frames by median combining and subtracted from object frames. In Optical photometric reduction, bias-correction, flat-fielding, and removal of cosmic rays were done on raw images to clean the science frames. The aperture photometry was carried out with these processed images using APPHOT package of IRAF. The zero-points of photometry were determined using the standard stars. 

The spectroscopic analysis was done using APALL task of IRAF. The TIRSPEC data was reduced with TIRSPEC pipe-line\footnote{\url{https://github.com/indiajoe/TIRSPEC/wiki}} (Ninan et al. 2014), and is cross-checked with the IRAF reduction. Both techniques agree well. The wavelength calibration in NICMOS-3 data was performed using OH skylines, while Argon lamp is used for TIRSPEC data. The wavelength calibration of optical spectra (HFOSC data) was done by using a FeNe arc spectrum. The science frames are divided by a standard star, observed at the similar airmass as science target, to remove the telluric features of the Earth's atmosphere. Then the flux calibration of the target stars is performed using the standard star. 
\section{Result and discussion}

\subsection{Optical Light Curves and Period}

\begin{figure*}
    \includegraphics[scale=0.60]{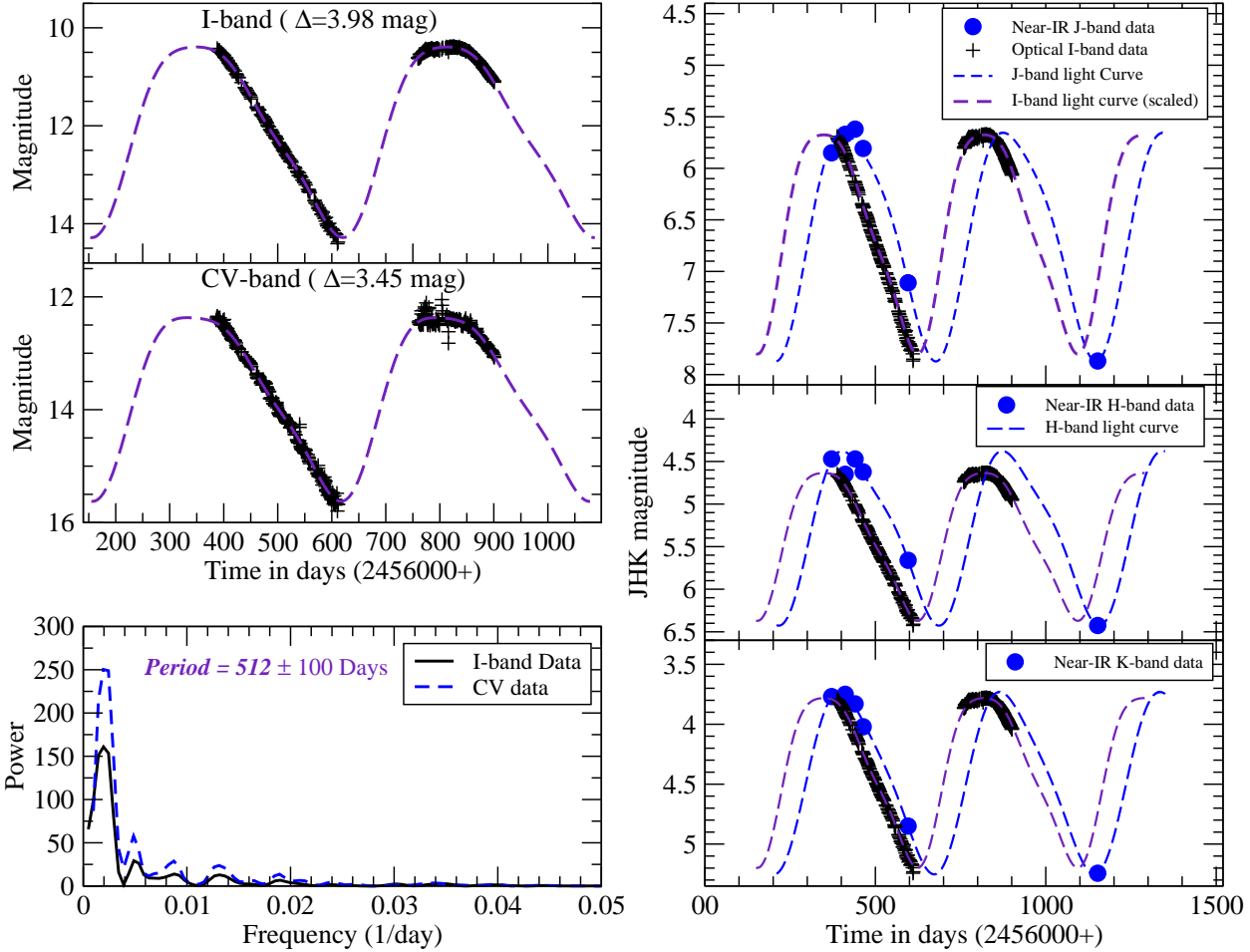}
    \caption{The left figures show the optical light curves of J2124+32 with fitting using Eqn.~1 in I-band (left top) and unfiltered CCD in 400--900 $nm$ (C) (left middle). The periodograms of optical light curves are shown in the left bottom panel. The NIR JHK light curves of J2124+32 are shown in the right three panels (JHK$^\prime$), where the filled circles are our observed NIR data points, while the solid lines are fitted light curve with P= 465 days. The optical I-band light curve (scaled with NIR light curves) is over-plotted on NIR light curves for comparison.}
    \label{fig1:lc}
\end{figure*}

Fig.~\ref{fig1:lc} shows the optical light curves in I-band and unfiltered (400--900~$nm$) CCD (Clear, C). The amplitudes of optical variability are estimated to be $\sim$ 4.00 mag, and $\sim$ 3.4 mag in I band and unfiltered CCD respectively. Such large amplitude of variability with a long-term period is only observed in a case of Mira-like variables \citep{Whitelock2003}. The general criteria to be a Mira variable, the amplitude variability in I-band should be greater than 1.0 mag \citep{Soszynski2012}.
  
For determination of the period of this object, we used the Lomb-Scargle (LS) periodogram \citep{Lomb1976, Scargle1982}, the algorithm publicly available at the starlink\footnote{\url{http://starlink.eao.hawaii.edu/starlink}} software database. The LS method is used to find out significant periodicity even with unevenly sampled data and verified successfully in several cases to determine periods from such sparse data sets \citep{Mondal2010}. The left bottom panel of Fig.~\ref{fig1:lc} shows the periodograms of the light curves determined from the LS method. We found the period to be 512 $\pm$ 100 days, and significant uncertainty in period is due to limited time coverage, which is not cover a complete periodic cycle. The period is also verified from PERIOD04\footnote{\url{http://www.univie.ac.at/tops/Period04}} \citep{Lenz2005}, which provides the same result.  

As noted in \citep{Lebzelter1999}, the classical method for deriving a period using Fourier analysis like LS method do not always provide satisfying results for light curves of AGB variables. Alternatively, we have used Fourier decomposition technique with the fitting of the following function as in \citet{Ngeow2013}, 
 
\begin{equation}
m~(t) = A_0 + \sum_{\kappa=1}^{N}  A_\kappa sin(\kappa \omega t + \phi _\kappa)
\end{equation} 
Where $\omega = 2\pi /P$, P is the period in days, $A_\kappa$ and $\phi _\kappa$ represent the amplitude and phase$-$shift for $\kappa ^{th}$ $-$order respectively, and N is the order of the fit. To fit the light curve, we use up to third order terms. From $\chi^2$ minimization technique, we find the best fitted period of 465$\pm$30 days on our light curves. The solid black line is the Fourier fit which is shown in Fig.~\ref{fig1:lc} along with observed data in the respective panels. The estimated period of this red object is consistent with red LMC Miras (J$-$K $\geq$ 2.08) having periods of 300$-$500 days \citep{Ita2004}.

\subsection{Near-Infrared light curves}

NIR $JHK^\prime$-bands photometric observations were carried out during 20 March 2013 to 09 May 2015 in a sparse sampling of the Mira cycle, and $JHK^\prime$ magnitudes are listed in Table~\ref{tab:phot}. Our $JHK^\prime$ magnitudes on first two epoch are reported immediately at ATEL \citep{Mondal2013}. The NIR $JHK^\prime$ light curves are shown in the right panels of Fig.~\ref{fig1:lc}. For comparison, the optical I-band light curve is also overplotted on those NIR light curves after scaling with $JHK^\prime$ magnitudes. The light curve is fitted to the Eqn.~1 mentioned above with the same period of the optical light curve fit. The amplitudes (peak to peak) of the variability are estimated from our light curves to be $\Delta J$ $\sim$ 2.2 mag, $\Delta H$ $\sim$ 1.9 mag and $\Delta K$ $\sim$ 1.5 mag respectively. These optical/NIR light curves confirm typical Mira behavior of strong wavelength dependence, i.e., the pulsation amplitude decreases with increasing wavelength \citep{Smith2002}. \citet{Bessell1996} showed that colors such as V-K change much more with phase than near-IR colors (e.g., (J-H),(H-K), (J-K)), which is another way of saying that visual amplitudes are larger than near-IR amplitudes.  The fact that the pulsation amplitude decreases with increasing wavelength is a result of the changes in $T_{eff}$ that accompany the changes in luminosity, and is exacerbated by the dependence of TiO opacities on temperature. Another interesting feature, a phase lag of about 60 days, corresponding to $\sim$0.13 of phase, between the optical and NIR maxima or minima is observed here in Fig.~\ref{fig1:lc}, which is seen in Mira variables \citep{Smith2002}. Such phase lag in oxygen-rich Miras may be due to the opacity of TiO molecules in their atmosphere \citep{Smith2006}, and even large visual amplitude might be due to formation and destruction of TiO molecules during the passage of periodic shock waves \citep{Reid2002}. The 2MASS (J-K) colour is 3.0 mag \citep{Cutri2003}. We also find a large (J-K) color index, ranging from 1.78 to 3.0 mag (Table~\ref{tab:phot}), which are again consistent with the source being an extreme red Mira variable \citep{Ita2004, Whitelock2000}. We rule out the object is  a semi-regular variables as the latter are not this red.

\subsection{Distances and Luminosity}

To estimate the distance to the source, we have used the period$-$luminosity (PL) relation for the Miras based on the distance modulus of the Large Magellanic Cloud (LMC) to be 18.50$\pm$ 0.02, the PL relation is taken from \citet{Ita2011} expressed as

\begin{equation}
 M_K = (-3.675\pm 0.076) \log P + (1.456\pm 0.173)
\end{equation}

According to the relation, the absolute K-band magnitude of the source is estimated to be  $M_K$ = $-$8.34$\pm$ 0.34 mag. We have also examined other available PL relations in \citet{Feast1989} and \citet{Whitelock2008}, and the value of $M_K$ is within uncertainty limit. The Galactic interstellar extinction in the direction of the source is $A_V$=0.57 ($A_K$=0.05; \citealt{Schlafly2011}) or $A_V$=0.68 ($A_K$=0.06; \citealt{Schlegel1998}). Taking $A_K$=0.05, we have obtained the distance (d) to the object, 3.27$\pm$0.02 kpc using the relation $m_k - M_K = 5\log d -5 + A_K$. The uncertainties in the distance measurement come from the estimated period, PL relation and photometric error of calculating K-band magnitude. We have also estimated bolometric magnitude $M_{bol}$ using the following PL relation for Miras \citep{Hughes1990}.

\begin{equation}
 M_{bol} = -3.22 - 7.76[\log(P/day) - 2.4] \pm 0.38
\end{equation}

We have estimated the bolometric magnitude of $M_{bol}$ = $-$5.29 $\pm$ 0.38 corresponding to luminosity $\sim$10250 L$_\odot$, which is consistent with other PL relation of \citet{Feast1989}. The O-rich Miras with P $\geq$ 420 d are over luminous \citep{Feast1989, Hughes1990}, which is consistent with our result. \citet{Feast1989} showed that for few Miras which have periods $>$ 420d are 0.7 mag brighter than expected extrapolation of PL relation. Also, \citet{Whitelock2003} suggested that Miras with periods $>$ 400 d have higher luminosities due to hot bottom burning. Putting our object on the Mass-Luminosity relation Fig.~6 of \citet{Hughes1990}, the object with P=465d and $M_{bol}$=$-$5.29 lies between the mass (M) range of 1.0$M_\odot$ $-$ 1.5$M_\odot$.

To understand the galactic location of the new mira variable,  we have estimated the scale height of the object from the distance above the galactic plane (Z)  following \citet{Jura1992},. We find that the scale height of the object is about 270 pc which agree with the thin disk population as given by \citet{Habing1988, Jura1992, Eggen1998, Juric2008}.

\subsection{Spectral Energy Distribution}
\begin{figure*}
\center
    \includegraphics[scale=0.40]{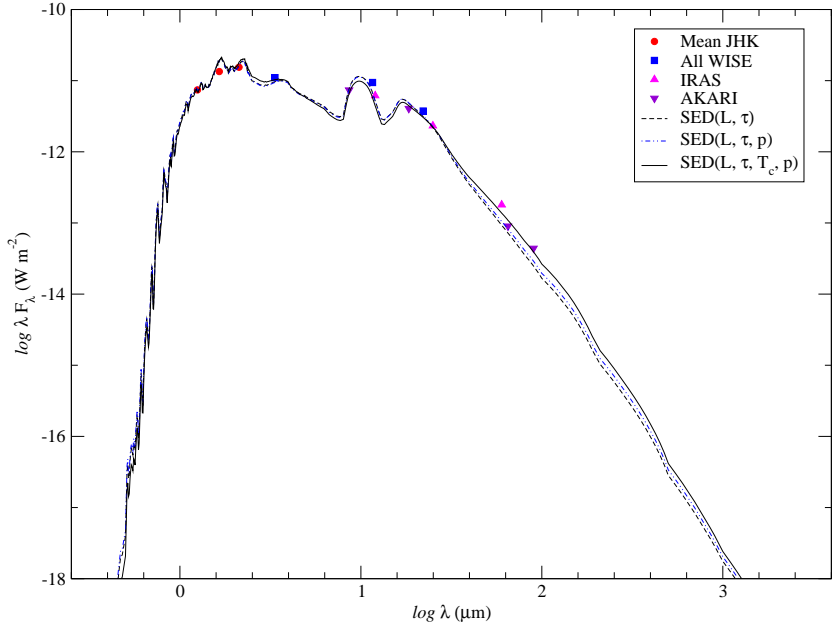}
  \caption{The SED of the target is shown here using multiwavelength data from NIR to far-IR, while the insets show different data source, e.g., our mean JHK$^\prime$, All WISE, AKARI, IRAS data. The SED (L, $\tau$) means fitted SED with L, $\tau$ as variables and $T_c$, p fixed. Similarly for  SED(L, $\tau$, p) and SED(L, $\tau$, $T_c$, p).}
 \label{fig2:sed}
\end{figure*}

We use near to far infrared multi-wavelength photometric data to generate the spectral energy distribution (SED) of the source J2124+32. The fluxes, used here for the SED fit, are taken from our JHK$^\prime$ measurements and archival mid to far Infrared catalogs. The mean JHK$^\prime$ are estimated from the NIR light curves. The mid-IR 3.35, 11.6, and 22.1 $\mu$m data are taken from AllWISE Data Release \citep{Cutri2013}; 8.61, 18.4 $\mu$m from AKARI/IRC all-sky Survey (ISAS/JAXA, 2010) \citep{Ishihara2010}; 12, 25 $\mu$m from IRAS Faint Source Catalog (IPAC 1992) and Far-IR 65, 90 $\mu$m data from AKARI/FIS; 60 $\mu$m from IRAS. The observed JHK$^\prime$ magnitudes are converted to flux densities using the zero magnitudes flux from \citep{Bessell1998}. 
 
The radiative transfer code, More of Dusty\footnote{\url{http://homepage.oma.be/marting/codes.html}} (MoD), developed by \citep{Groenewegen2012}, was used to model the dusty circumstellar shell of Mira variable. The MoD is a modified version of a publicly available 1D dust radiative transfer code, ``DUSTY" version 2.01 \citep{Ivezic1999}. The detail mathematical formulation of ``DUSTY" was described elsewhere \citep{Ivezic1997}. The MoD works to find best-fitted parameters, e.g., luminosity (L), dust optical depth ($\tau_{0.55}$), dust temperature at the inner radius ($T_c$) and slope of the density law (p), etc. in the minimization process. The quality of the fit is obtained through a $\chi^2$ analysis.

The interstellar reddening ($A_V$), hydrostatic model atmosphere, the effective temperature of the model atmosphere, distance (D), outer radius in a unit of inner radius, dust properties, etc. are provided as an input in the master-input file. The adopted dust species are taken as mixture of Mg$_{0.8}$Fe$_{1.2}$SiO$_4$:AlO:Fe=100:0:5, calculated using the distribution of hollow spheres (DHS) with a mean grain size a= 0.15 $\mu$m and a maximum volume fraction of a vacuum core $f_{max}$=0.7 $\mu$m (private communication with M. A. T. Groenewegen). We adopted such dust composition because our spectroscopic studies concluded that the star is O-rich as will be discussed in subsequent sections. The MARCS\footnote{\url{http://marcs.astro.uu.se/}} hydrostatic model atmospheres \citep{Gustafsson2008} were used for the spectra of the central stars. The outer radius is set to 1600 times the inner radius, where dust temperature becomes approximately 20 K. In the model, the distance (d) of 3.27 kpc and $A_V$ were adopted as an input to the standard model as discussed in the previous section. The L, $\tau$, $T_c$ and p could be fitted or set to a fixed value.

\begin{table*}
\begin{center}
\caption{MoD Models for J2124+32 for three different cases. }
\label{tab:sed}
%\resizebox{1.0\textwidth}{!}{
\begin{tabular}{lcccccr}
\hline
\hline
Star & L (L$\odot$) & $\tau_{0.55}$  & $T_c$ [K]& p & $\chi^2_{red}$ \footnote{$\chi^2_{red}$ = reduced $\chi^2$ for the fit of photometric data} & Remarks \\
\hline
    & 9331 $\pm$ 1545 & 10.064 $\pm$ 2.489 & 1000 fixed & 2.0 fixed & 819 & $T_c$, p fixed  \\
J2124+32 & 9255 $\pm$ 1659 &9.913 $\pm$ 2.746 & 1000 fixed  & 1.95 $\pm$ 0.29 & 818 & $T_c$ fixed  \\
   & 9282 $\pm$ 1590 & 11.178 $\pm$ 3.975 & 1248 $\pm$ 511 & 1.75 $\pm$ 0.28 & 806 & All free\\

\hline
\hline
\end{tabular}
\end{center}
\end{table*}

To get the best fit of photometric data, we generated multiple SEDs considering MARCS model atmospheres of temperature range 2600-3600 K in step of 200 K and setting  other free parameters (e.g., L, $\tau_{0.55}$, T$_c$, p) either fixed or variable as listed in Table~\ref{tab:sed}. The model provides minimum reduced $\chi^2$ value at MARCS model atmospheres of temperature 2800 K, and the parameters of MoD fitting at that temperature for different cases are given in Table~\ref{tab:sed}. Our best fit SED of all photometric measurements for various cases are shown in Fig.~\ref{fig2:sed}. We got the minimum value of reduced $\chi^2$, when all the parameters set free. The $\chi^2$ of the fit is typically large as we include significant JHK$^\prime$ variability in our SED fit, and non-simultaneous taking of the photometric data. The provided errors are therefore internal errors scaled to a reduced $\chi^2$ of 1 (private communication with M. A. T. Groenewegen). The best-fitted SED provides L= 9282 $\pm$ 1590 L$_\odot$ that is comparable to our P-L based estimation in the earlier section, $\tau$= 11.178 $\pm$ 3.975, $T_c$= 1248 $\pm$ 511 K and p= 1.75 $\pm$ 0.28. The best-fit parameters, dust properties and actual grain size translate into a current mass-loss rate of $0.7\times10^{-6} M_\odot yr^{-1}$ by assuming a dust-to-gas ratio of 0.005 and an expansion velocity of 10 km/s.

\subsection{Optical/NIR spectroscopic studies}

\begin{figure*}[h!]

\includegraphics[scale=0.75, clip=true]{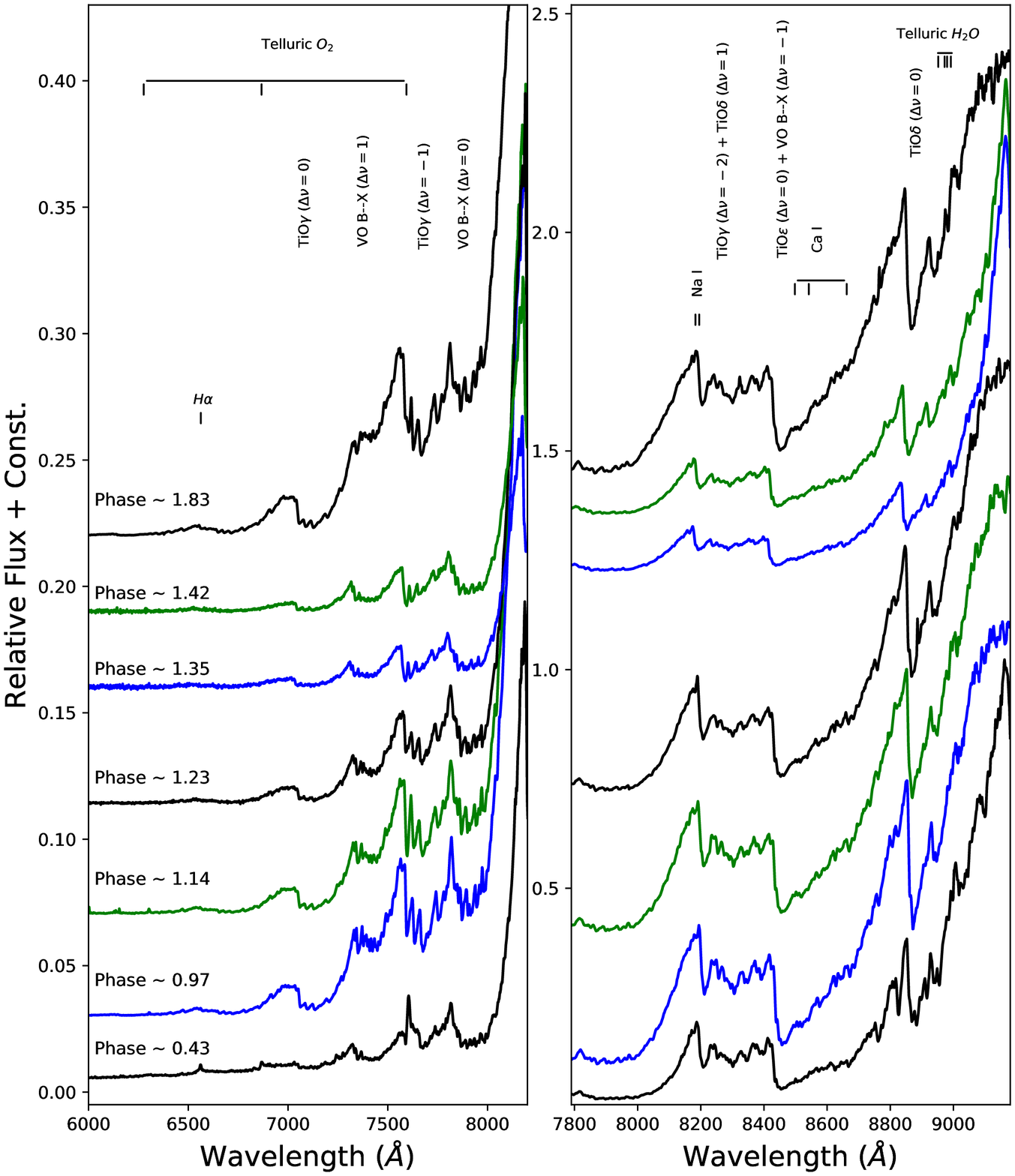}
\caption{In top two panels, the optical spectra of the object in the range 6000-9200 \AA,~ which show visible features of TiO and VO bands at different phases of the Mira including H$_\alpha$ emission at phase 0.43. The spectra have been normalized to unity at 9165 \AA,~ and offset by constant values 0.0, 0.03, 0.07, 0.11, 0.16, 0.19, 0.22 (left panel) and 0.0, 0.06, 0.36, 0.70, 1.20, 1.35, 1.40 (right panel) respectively with respect to the bottom-most spectra. } 
\label{fig3:optspectra}
\end{figure*}
 
\begin{figure*}[h!]

\includegraphics[scale=0.70, clip=true]{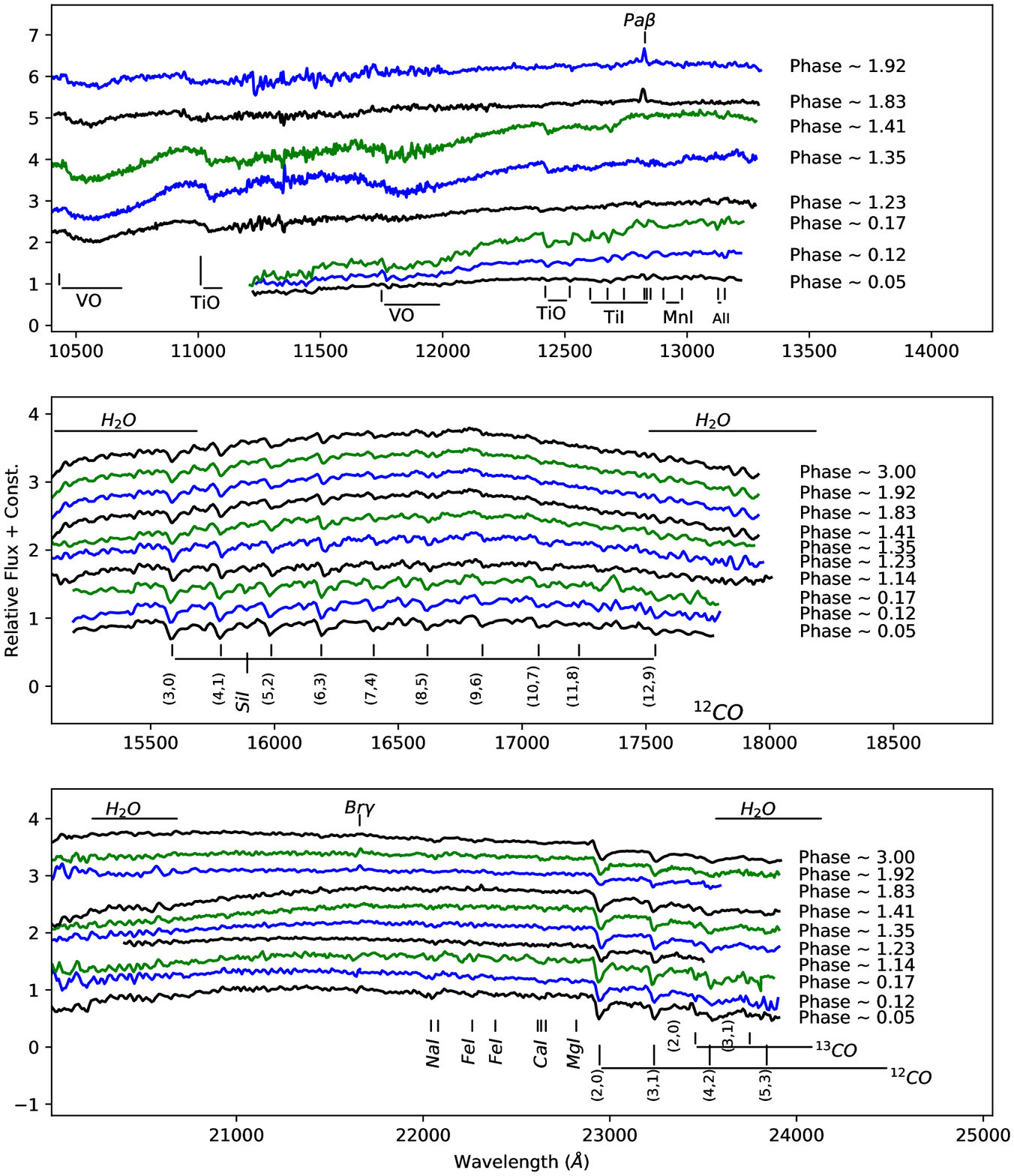}
\caption{ The NIR JHK-band spectra in the wavelength range 1.02$-$2.39 $\mu$m at eight different phases of the Mira are shown here in the $1^{st}$, $2^{nd}$, and $3^{rd}$ panels, respectively. First three NIR spectra from the bottom (phase $\sim$0.05, 0.12, 0.17) are taken with the NICMOS-3 instrument on 1.2m Mt.Abu telescope, and rest are observed with TIRSPEC instrument on 2.0m HCT. In J-band, molecular bands like TiO, VO, and few atomic lines are present in the spectra. The Pa$\beta$ emission line appears at two phases (1.83 and 1.92). The H-band spectra in the wavelength range 1.52$-$1.80 $\mu$m show strong four $^{12}$CO second overtone bands including several OH lines. In the K-band spectra, the $^{12}$CO first overtone bands are dominated features in the spectra, and Na I and Ca I are seen at 2.20 $\mu$m and 2.26 $\mu$m, respectively. The Br$\gamma$ (at 2.16 $\mu$m) emission line appears at phases (1.83, 1.92 and likely at 0.05). The spectra have been normalized to unity at 12000 \AA (J-band), 16500 \AA (H-band), 21700 \AA (K-band), and offset by constant values 0.0, 0.30, 0.60, 0.11, 1.70, 2.40, 3.30, 4.20, 5.10 ($1^{st}$ panel) respectively with respect to the bottom-most spectra of the same panel (J-band), and 0.30 to each spectrum of H and K-band).}   
\label{fig4:jhkspectra}
\end{figure*}
 
Multi-epoch optical/NIR spectroscopic studies on Mira variables at different variability phases probe the dynamic stellar atmosphere and help us also to understand the pulsational related variations of fundamental parameters. The visual spectra of the object taken at several variability phases are shown in Fig.~\ref{fig3:optspectra}. The optical spectra are dominated by molecular absorption bands of TiO and VO in the wavelength range of 7000 to 9000 \AA ~like red M stars \citep{Castelaz2000, Rayner2009, Bessell1989,  Fluks1994}. Visual comparison shows that all the optical spectra of the MASTER OT object correspond to O-rich spectral types later than M7.

The NIR spectra were taken immediately after transient alert of \citet{Tiurina2013} and continued over several variability phases as listed in Table~\ref{tab:log}. The JHK spectra are shown in Fig.~\ref{fig4:jhkspectra}. Molecular features of TiO and VO bands dominate in the J-band spectra as shown in the 1st panel of Fig.~\ref{fig4:jhkspectra}. But no ZrO twin features at 1.03, and 1.06 $\mu$m is seen here, which is a primary indicator for S-type Mira stars \citep{Hinkle1989, Wright2009}. The most prominent feature of VO band is seen in one of our spectra, which covers from 1.02 $\mu$m, which signifies it's oxygen-rich M$-$type star \citep{Wright2009}.

In the 2nd panel of Fig.~\ref{fig4:jhkspectra}, the H-band spectra show the $^{12}$CO second overtone series at 1.5582, 1.5780, 1.5982, 1.6189, 1.6397, 1.6610, 1.6840, 1.7067 $\mu$m including OH molecular bands all over the band \citep{Rayner2009}. These second overtone series develop at the objects having a photospheric temperature of $\leq$ 5000 K, and their strength is dependent on the photospheric temperatures \citep{Lancon2007}.  The OH molecular bands are prominent in M-type, very weak in S-type and absent in carbon stars as all oxygen is locked up in CO. No C$_2$ band at 1.77 $\mu$m is seen here as observed in carbon-rich Miras. 

The K band spectra are shown in the most bottom panel of Fig.~\ref{fig4:jhkspectra}. The $^{12}$CO first overtone series at 2.2935, 2.3227 $\mu$m, neutral atomic lines of  Na I doublet at 2.20 $\mu$m, Ca I triplet at 2.26 $\mu$m are seen in our spectra, which are original features in M-type evolved stars \citep{Rayner2009}. The $^{12}CO$(2,0) first overtone band heads are the strongest absorption feature in K band in cool stars, and its strength depends on both luminosity and effective temperature \citep{Ramirez1997, Cesetti2013}. The first overtone of $^{13}$CO is also visible in our spectra, though its strength is relatively weak compared to $^{12}$CO first overtone series.  

The H$\alpha$ emission line is seen only at phase $\sim$0.43. The Pa$\beta$ emission line appears in two visual phase 1.83 and 1.92. The Br$\gamma$ appears at  phase 1.83 and reach maximum intensity at 1.92. These emission lines appear in the spectra due to shock wave generation in the Mira atmosphere. It is to be noted that H$\alpha$ is seen at phase 0.43, while Pa$\beta$ and Br$\gamma$ are seen near maximum phase. While other studies find Balmer lines usually around maximum phase (e.g., \citealt{Fox1984, Fox1985}).

The overall low-resolution continuum shape of J, H and K spectra turn downwards at the end of the band indicating absorption due to broad H$_2$O absorption features centered at 1.4 $\mu$m, 1.9 $\mu$m \& 2.7 $\mu$m in our spectral coverage \citep{Rayner2009}. It is strongly visible that the continuum shape is changing over the variability phases.
  
\subsubsection{Phase dependent spectral variability}

\begin{figure*}
\center
\includegraphics[scale=0.60,clip=true]{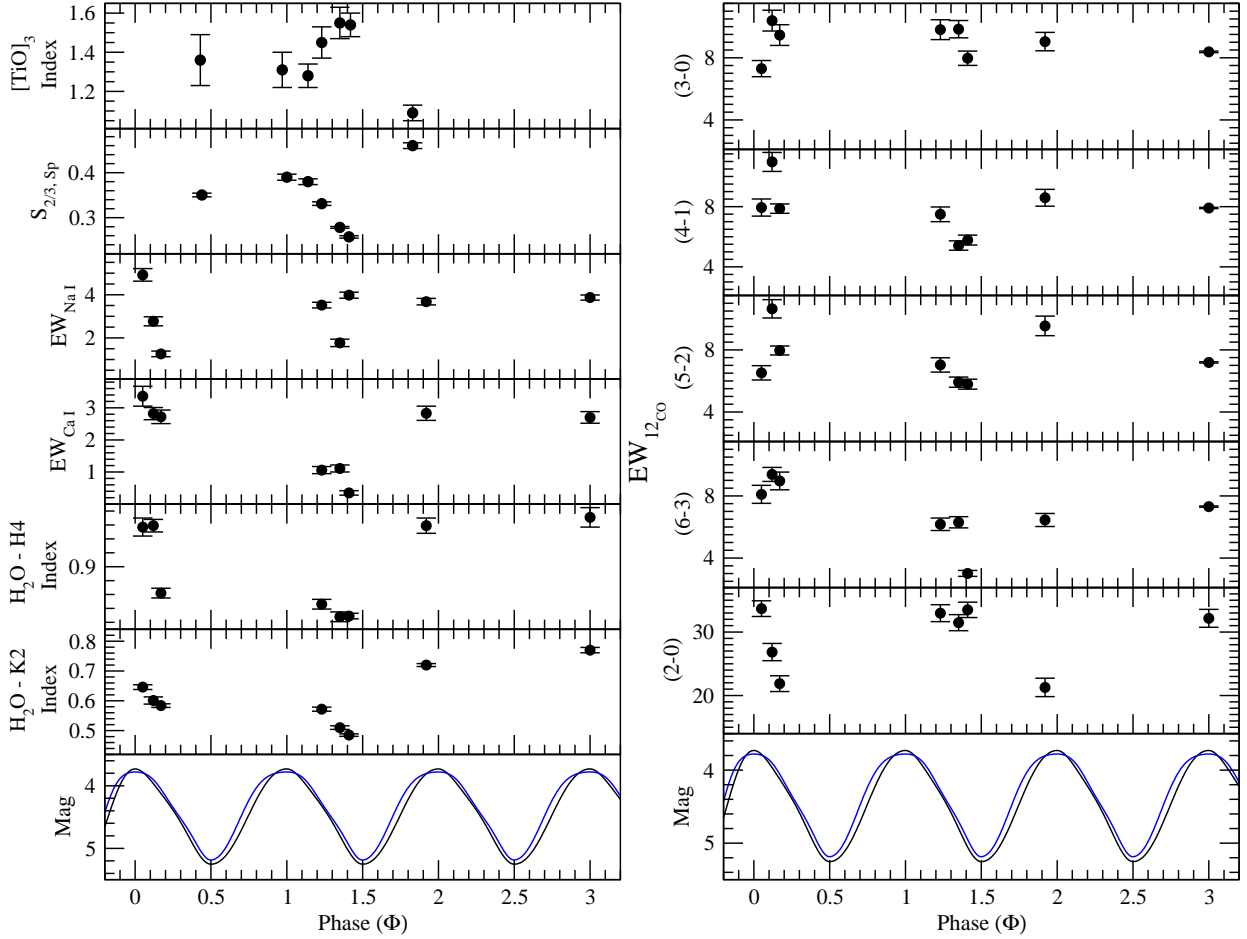} 
\caption{The phase variation of [TiO]$_3$, S$_{2/3, Sp}$, Na I, Ca I, H$_2$O-K2 equivalent width/index are shown with the visual phase. The bottom panel shows the K-light curve (black) and I-light curve (scaled with K-light curve).}
\label{fig5:SiNaCa}
\end{figure*}  

The variation in the absorption depth of TiO and VO with $T_{eff}$ is well documented from long before \citep{Merrill1962}. \citet{Lockwood1972} found that TiO band-strength indices decrease in stars later than M7, and it saturates in M9 Miras.  We explore the phase variation of [TiO]$_2$ index centered at 7100 \AA~as described in \citet{Connell1973}. But, [TiO]$_2$ index shows no significant variation over the phase. It is to be noted that we consider here the optical phase variation.

Another triple-headed absorption bands of TiO at 8433, 8442, 8452 \AA,~[TiO]$_3$, are considered following \citet{Zhu1999} for such phase variation studies. The [TiO]$_3$ is defined as,

\begin{equation}
[TiO]_3 = -2.5 log(\frac{F_\lambda}{F_C})
\end{equation}
where, $F_\lambda$ is flux at $\lambda$ and $F_C$ is the interpolated continuum at $\lambda$. The pseudo continuum was generated in the left window of the band at 8390 $-$ 8410 \AA,~and right windows at 8700 $-$ 8725 \AA.~Then the spectrum is normalized by the continuum, and the $\frac{F_\lambda}{F_C}$ is measured in the range 8455 $-$ 8470 \AA. ~As shown in Fig.~\ref{fig5:SiNaCa}, the [TiO]$_3$ index shows significant variation with the pulsation phase, it increases as the visual brightness decreases, becomes constant at certain phases (saturation effect), and then further decreases with visual brightness increases.

The Fig.~\ref{fig3:optspectra} shows three strong absorption bands 706-724 nm (band1), 770-807 nm (band2) and 829-857 nm (band3),  which are related to spectral types of static giants  \citep{Fluks1994}. The flux ratio at different bands is defined by \citet{Fluks1994}. We studied the flux ratio at those bands and found that the  band2 to band3 ratio (as S$_{2/3, Sp}$) show pronounce variation with phase as shown in the Fig.~\ref{fig5:SiNaCa}. The S$_{2/3, Sp}$ becomes maximum at visual maximum while minimum at a visual minimum.

We have measured the equivalent widths (EWs) for the atomic spectral  Na I at 2.206 $\mu$m, Ca I at 2.263 $\mu$m. The continuum bands for Na I  and Ca I are taken from \citet{Ramirez1997} and mentioned in Table~\ref{tab:featureband}. The spectra were normalized by the local pseudo-continuum, and  EWs of the particular atomic features were estimated by using splot task in the IRAF by fitting a Gaussian function to each feature.  The Na doublet is present in the 2.2051 -2.2099 $\mu$m region but blended with metallic lines like Si I (2.2069 $\mu$m), Sc I (2.2058 and 2.2071 $\mu$m) and V I (2.2097 $\mu$m) in our low-resolution spectra. The Ca triplet appear in the 2.2609-2.2665 $\mu$m regions, and is very sensitive to temperature. The 3rd and 4th panels of Fig.~\ref{fig5:SiNaCa} show their phase variation. The atomic lines, in particular, are weak features that are very difficult to measure in low-resolution spectra,
and no significant trends with phase are apparent, except possibly for Ca. 

\begin{figure*}
\center
	\includegraphics[scale=0.55]{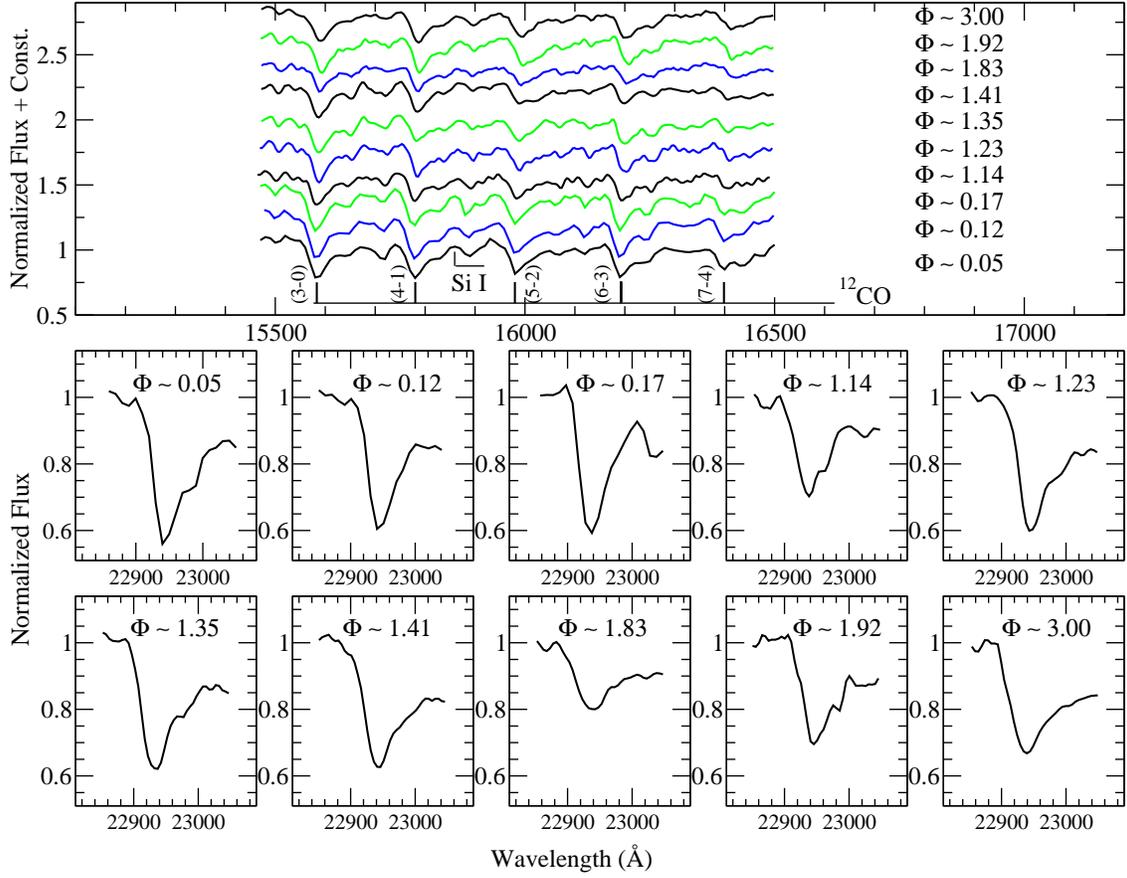}
\caption{The changing shape of CO-second overtones and one CO first overtone at 2.29$\mu$m with phases are shown here} 
\label{fig6:CO2}
\end{figure*}

In the 1.5-2.4 $\mu$m region, the first-overtone ($\Delta\nu$=2) and the CO second overtone($\Delta\nu$=3) band heads of CO are the dominant features in the spectra as mentioned in the earlier section. We studied the phase dependent variations of these 2-0 (first overtone) and 3-0, 4-1, 5-2, 6-3 (second overtone) $^{12}$CO band heads. We estimated the EWs of CO at different phase considering the local continuum as mentioned in the Table~\ref{tab:featureband}, the continuum has been fitted with first order spline fitting function (linear interpolation) at peak points.  We estimated the EWs of CO at different phase as shown in Fig.~\ref{fig5:SiNaCa}. The EWs of 3-0 and 4-1 band heads doesn't demonstrate any significant change during the pulsation cycle. The EWs of 2-0 first overtone band change significantly.  The EWs of 5-2 and 6-3 band heads showed weak variation. The result should be taken as a caution as such small variation might occur due to the computational artifact, continuum selection and blending effect from the weak OH-lines. The shock wave, in general, propagates through the Mira atmosphere (CO first overtone forming-layer) in between optical phase $\sim$ 0.1-0.2, which alters velocity profile known from high-resolution spectra \citep{Hinkle1979b, Nowotny2010}. Even in our low to intermediate resolution spectra, the CO absorption features appear to change in Fig.~\ref{fig6:CO2}. It is expected that the combination of modified individual lines will also modify the shape of the band as a whole.
   
The shape of NIR spectra in Miras is dominated by strong, and broad water bands centered at 1.4 $\mu$m and 1.9 $\mu$m and 2.7 $\mu$m regions \citep{Johnson1968}. The depth of water bands vary with Mira phases and become strongest at minimum light \citep{Strecker1978}.  Due to this water absorption, the H and K-band spectra bend downward at the end of both bands. So, the curvature of the spectral changes depending on the water absorption. To quantify water absorption in our data, we measured the H$_2$O-H4 and H$_2$O-K2 indices, which shows the curvature variation of the spectra. The H$_2$O$-$K2 Index is taken from \citet{Rojas2012} as,

\begin{equation}
H_2O-K2 = \frac{\langle F (2.070-2.090)\rangle / \langle F (2.235-2.255)\rangle}{\langle F (2.235-2.255)\rangle / \langle F (2.360-2.380)\rangle}
\end{equation}

Where $\langle F (a-b)\rangle$ represents the median flux level in the wavelength range defined by a and b in $\mu$m. \citet{Rojas2012} calculated this index for M and K-type dwarfs. Here that same index is explored using the spectra for M-type giants in the wavelength range of 2.07$\mu$m to 2.38 $\mu$m. 

 For the H-band spectra, several authors defined spectro-photometric indices to measure water absorption (e.g., \citealt{Allers2007, Weights2009, Scholz2012}). Here we define a new H$_2$O-H4 index as,

\begin{equation}
H_2O-H4 = \frac{\langle F (1.531-1.541)\rangle / \langle F (1.670-1.690)\rangle}{\langle F (1.670-1.690)\rangle / \langle F (1.742-1.752)\rangle}
\end{equation}
where $\langle F (a-b)\rangle$ represents the median flux level in the wavelength range defined by a and b in $\mu$m.   

In Fig.~\ref{fig5:SiNaCa}, the phase variation of these two H$_2$O-H4 and H$_2$O-K2 indices are shown. The smaller value of H$_2$O indices in H and K corresponds to greater amounts of H$_2$O opacity. The signification variations of these indices apparent in the Fig.~\ref{fig5:SiNaCa}. In Fig.~\ref{fig5:SiNaCa}, we see that H$_2$O-H4 and H$_2$O-K2 indices show significant variation with the pulsation cycles, the value of indices are strongest at visual maximum while weak at a visual minimum i.e. H$_2$O opacity is stronger at visual minimum. Our result confirms the trends already seen by \citet{Strecker1978}. 

\begin{table*}
\begin{center}
\caption{Definitions of Spectral Bands }
\label{tab:featureband}
%\resizebox{1.0\textwidth}{!}{
\begin{tabular}{cccc}
\hline
\hline
Feature & Bandpass ($\mu$m) & Continuum bandpass ($\mu$m) &   Ref \\
\hline
$[TiO]_3$ & 0.8455-0.8725 &  0.8390-0.8410, 0.8700-0.8725 & 1  \\
Na I	    & 2.204-2.211   & 2.191-2.197, 2.213-2.217      & 2    \\
Ca I	    &2.258-2.269    &   2.245-2.256, 2.270-2.272    & 2   \\
$^{12}CO$(3-0)  & 1.5550-1.5625 & -& 2 \\
$^{12}CO$(4-1)  & 1.5752-1.5812 &- & 2 \\
$^{12}CO$(5-2)  & 1.5952-1.6020 &- & 2 \\
$^{12}CO$(6-3)  & 1.6170-1.6220 &- & 2 \\
$^{12}CO$(2-0)  & 2.289-2.302 & 2.270-2.272, 2.275-2.278 & 2 \\ 
        &            &  2.282-2.286, 2.288-2.289    & \\
\hline
\hline
\end{tabular}
\end{center}
$^1$\citet{Zhu1999}; $^2$\citet{Ramirez1997}.
\end{table*}

\begin{table*}
\begin{rotatetable*}
\begin{center}
%\rotate
\caption{Phase Dependent Study }
\label{tab:parameters}
\resizebox{1.1\textwidth}{!}{
%\begin{threeparttable}
\begin{tabular}{cccccccccccccccc}
\hline
\hline\\
 Date of  & Optical & $[TiO]_3$  & S$_{2/3, Sp}$ & Na I & CaI & CO & CO & CO & CO & CO  & $H_2O$-H4 & $H_2O$-K2  & Sp.   \\
%\cline{3-12}
Obs. & Phase & Index  & & 2.20 $\mu$m & 2.26 $\mu$m  & 3-0 & 4-1 & 5-2 & 6-3 & 2-0 & Index & Index &  $Type^{1}$   \\
\hline
2013 Mar 21.99 & 0.05 &- & - & 4.92$\pm$0.29	& 3.36$\pm$0.31 & 7.30$\pm$0.52 & 7.94$\pm$0.57 & 6.52$\pm$0.46 & 8.1$\pm$0.58 & 33.68$\pm$1.23 & 0.96$\pm$0.01 &  0.65$\pm$0.01  &-  \\
2013 Apr 29.99 & 0.12 &- & - & 2.77$\pm$0.21 	& 2.82$\pm$0.19 & 10.39$\pm$0.67 & 10.97$\pm$0.63 & 10.64$\pm$0.59 & 9.38$\pm$0.45 & 26.84$\pm$1.36 & 0.96$\pm$0.01 & 0.60$\pm$0.01  &-  \\
2013 May 30.95 & 0.17 &- &- & 1.26$\pm$0.13	& 2.72$\pm$0.21  & 9.46$\pm$0.67 & 7.87$\pm$0.31 & 7.96$\pm$0.29 & 8.96$\pm$0.57 & 21.87$\pm$1.24 & 0.86$\pm$0.01 & 0.58$\pm$0.01  &- \\
2013 Oct 15.57 & 0.44 & 1.36$\pm$0.13  &0.350 $\pm$ 0.004 &-&-&-&-&-&-&-&-&-& M9.5 \\
2014 May 21.91 & 1.0 & 1.31$\pm$0.09  &0.390 $\pm$0.007&-&-&-&-&-&-&-&-&-& M9 \\
2014 Aug 18.72 &1.14 &-  & - & 1.37$\pm$0.11   & 1.10$\pm$0.12 & 8.03$\pm$0.44 & 6.02$\pm$0.33 & 6.22$\pm$0.34  & 6.023$\pm$0.32 & 18.65$\pm$1.08 & 0.90$\pm$0.011 &  -  &- \\
2014 Aug 19.83 & 1.14 & 1.28$\pm$0.06  &0.380$\pm$0.006&-&-&-&-&-&-&-&-&-& M9 \\
2014 Oct 06.73 & 1.23 & 1.45$\pm$0.08 & 0.331 $\pm$0.003   & 3.52$\pm$0.13	& 1.06$\pm$0.11  & 9.81$\pm$0.64 & 7.49$\pm$0.49 & 7.03$\pm$0.46  & 6.173$\pm$0.40 & 32.97$\pm$1.33 & 0.85$\pm$0.01 & 0.57$\pm$0.01 & M9  \\
2014 Dec 12.55  & 1.35 & 1.55$\pm$0.08  & 0.278 $\pm$0.002  & 1.77$\pm$0.17 & 1.11$\pm$0.11  & 9.84$\pm$0.56 & 5.42$\pm$0.31 & 5.92$\pm$0.33  & 6.30$\pm$0.36 & 31.47$\pm$1.27 & 0.83$\pm$0.01 & 0.51$\pm$0.01 & M10 \\ 	
2015 Jan 13.58  & 1.41 &- & - & 3.98$\pm$0.14	& 0.35$\pm$0.07 & 7.97$\pm$0.46 & 5.78$\pm$0.33 & 5.788$\pm$0.32  & 3.01$\pm$0.19 & 33.49$\pm$1.21 & 0.83$\pm$0.01 & 0.48$\pm$0.01 &- \\
2015 Jan 18.56 & 1.41 & 1.54$\pm$0.06  & 0.257 $\pm$0.003&-&-&-&-&-&-&-&-&-& M10 \\
2015 July 05.79	& 1.83 & 1.09$\pm$0.04  & 0.460 $\pm$0.006 & 1.94$\pm$0.09  & 1.57$\pm$0.17  & 6.85$\pm$0.31 & 6.66$\pm$0.30 & 6.03$\pm$0.27  & 4.32$\pm$0.20 & 19.52$\pm$0.93 & 0.93$\pm$0.01 & -  & M8.5 \\
2015 Aug 11.88 & 1.92 &- & - & 3.68$\pm$0.15	& 2.83$\pm$0.22 & 9.04$\pm$0.59 & 8.59$\pm$0.56 & 9.54$\pm$0.63  & 6.45$\pm$0.42 & 21.27$\pm$1.45 & 0.96$\pm$0.01 & 0.72$\pm$0.01 &- \\
2016 Dec 19.54 & 3.00 &- & - & 3.87$\pm$0.12	& 2.70$\pm$0.18  & 8.38$\pm$0.05 & 7.91$\pm$0.04 & 7.19$\pm$0.04  & 7.31$\pm$0.04 & 32.15$\pm$1.41 & 0.97$\pm$0.01 & 0.77$\pm$0.01 &-\\	       
\hline
\hline
\end{tabular}
}
\end{center}
 $^{1}$The spectral type has been estimated using the correlation with $[TiO]_3$ Index
 \end{rotatetable*}
\end{table*}

\subsubsection{Spectral type}

The depth of triple-headed absorption band of TiO (8432, 8442 and 8452 \AA ~) at the optical spectra is excellent spectral type indicator. Following \citet{Zhu1999}, we estimate the spectral type (ST) of the object at different variability phase, which is defined as the $[TiO]_3$ index ([TiO] at 8450 \AA).~

\begin{equation}
 ST = 2.43 + 6.65[TiO]_3 - 1.12[TiO]_3^2
\end{equation}

We have measured [TiO]$_3$ index using Eqn.~4 from the spectra as described in the earlier section 3.5.1, the ST is estimated using the above relation in Eqn.~7, and the STs at different phases are shown the Table~\ref{tab:parameters}. The ST of the object varies from M8.5 to M10 over the phase of the pulsation cycle in our limited phase coverage. The saturation effect of [TiO]$_3$ is problematic to estimate the STs over phases as described in the earlier section 3.5.1.

\section{Summary and conclusion}

From long-term optical/NIR photometric and spectroscopic observations, we have characterized the MASTER OT J2124+32. Our main results are summarized as follows :

\begin{enumerate}
\item From the best-fit of optical/NIR light curves, we estimated the variability period of the object as 465 $\pm$ 30 days. The strong wavelength-dependent variability amplitudes in optical to NIR wavelengths are observed as $\Delta$I $\sim$4 mag, $\Delta$C (400--900~$nm$) $\sim$3.4 mag, $\Delta J$ $\sim$2.2 mag, $\Delta H$ $\sim$1.9 mag and $\Delta K$ $\sim$1.5 mag. Such large periods and strong wavelength-dependent variability amplitude are seen in Miras only.  Interestingly, a phase lag of $\sim$ 60 days between the optical and NIR light curves like a Mira variable is also seen. Large (J-K) NIR colors varying 1.78 --3.0 mag over phases, signifies that it is a red object like cool Miras.

\item From Period-Luminosity (PL) relation, the distance to the source is estimated as 3.27 $\pm$ 0.02 kpc. The absolute bolometric magnitude is determined as -5.29 $\pm$ 0.38 mag, corresponding to the luminosity of $\sim$10,250 L$_\odot$.

\item Using DUSTY based MoD code, we have done the SED fitting the NIR to Far-IR data. The best fit SED of all photometric measurements provides an effective temperature of 2800 K, and dust shell temperature 1248 K. The SED provides luminosity of the object 9282 L$_\odot$, which is comparable to the P-L based estimation, and mass-loss rate of $0.7\times10^{-6} M_\odot yr^{-1}$.

\item From Optical/NIR spectra, we find that the source has all kind spectral signatures of a cool M-type star. The spectral features indicate an O-rich Mira as it shows most prominent feature of VO band and TiO bands.  We rule out S- or C-type nature as ZrO bands at 1.03 and 1.06 $\mu$m  and  $C_2$ band head at 1.77 $\mu$m are absent.  

\item The phase-dependent of optical/NIR spectral features are studied. Notable variable features in all atomic and molecular lines (e.g., TiO, Ca I, H$_2$O and CO bands) over phases are seen here like commonly observed in Miras. Our optical spectral data show an apparent variation of the spectral type of the object over the pulsation cycle. 

\end{enumerate}

In conclusion, all these observational properties of the object J2124+32 confirms a new O-rich Mira variable toward the Cygnus.

\section*{Acknowledgements}
The authors are very much thankful to the anonymous referee for his/her critical and valuable comments, which help us to improve the paper. This research work is  supported by S N Bose National Centre for Basic Sciences under Department of Science and Technology, Govt. of India. The authors are thankful to the HTAC members and staff of HCT, operated by Indian Institute of Astrophysics (Bangalore); staff of the Mt. Abu observatory, operated by Physical Research Laboratory (Ahmedabad). SG is thankful to M. A. T. Groenewegen for helpful discussions and valuable suggestions on DUSTY code and MoD.

%%%%%%%%%%%%%%%%%%%%%%%%%%%%%%%%%%%%%%%%%%%%%%%%%%


\software{starlink \citep{Currie2014}, IRAF (Tody 1986, Tody 1993), TIRSPEC pipe-line \citep{Ninan2014}, PERIOD044 \citep{Lenz2005}, More of Dusty (MoD; \citealt{Groenewegen2012}), DUSTY \citep{Ivezic1997}, MARCS \citep{Gustafsson2008}}


\begin{thebibliography}{}
\bibitem[Allers et al.(2007)]{Allers2007} Allers, K. ~N., Jaffe, D. ~T., Luhman, K. ~L., et al.\ 2007, \apj, 657, 511
%\bibitem[Alvarez \& Plez(1998)]{Alvarez1998} Alvarez, R., Plez, B.,\ 1998, A\& A, 330, 1109
\bibitem[Alvarez et al.(2000)]{Alvarez2000} Alvarez, R., Jorissen, A., Plez, B., et al.\ 2000, A\& A, 362, 655
\bibitem[Aringer et al.(2009)]{Aringer2009} Aringer, B., Girardi, L., Nowotny, W., et al.\ 2009, A\& A, 503, 913
\bibitem[Bessell et al.(1989)]{Bessell1989} Bessell, M. S., Brett, J. M., Wood, P. R., Scholz, M.,\ 1989, A\&A, 213, 209
\bibitem[Bessell, Scholz \& Wood(1996)]{Bessell1996} Bessell, M. S., Scholz, M., Wood, P. R.,\ 1996, A\&A, 307, 481
\bibitem[Bessell et al.(1998)]{Bessell1998} Bessell, M. S., Castelli, F., Plez, B.,\ 1998, A\& A, 333, 231
\bibitem[Bieging et al.(2002)]{Bieging2002} Bieging et al.\ 2002, A\&A 384, 965
%\bibitem[Bowen(1988)]{Bowen1988} Bowen, G. H.,\ 1988, \apj, 329, 299
%\bibitem[Brett(1990)]{Brett1990} Brett, J. M.,\ 1990, A\&A, 231, 440
\bibitem[Castelaz \& Luttermoser(1997)]{Castelaz1997} Castelaz, M. W. \& Luttermoser, D. ~G.\ 1997, AJ, 114, 1584
\bibitem[Castelaz  et al.(2000)]{Castelaz2000} Castelaz, M. W. et al.,\ 2000, AJ, 120, 2627
\bibitem[Cesetti et al.(2013)]{Cesetti2013} Cesetti, M., Pizzella, A., Ivanov, V. D., et al.\ 2013, A\&A, 549, 129
\bibitem[Currie et al.(2014)]{Currie2014} Currie, M. J.; Berry, D. S.; Jenness, T.; Gibb, A. G.; Bell, G. S.; Draper, P. W.,\ 2014, ASPC, 485, 391
\bibitem[Cutri et al.(2003)]{Cutri2003} Cutri, R. M., et al.\ 2003, yCat, 2246, 0
\bibitem[Cutri et al.(2013)]{Cutri2013} Cutri, R. M., et al.\ 2013, yCat, 2328, 0
\bibitem[Dejonghe \& van Caelenberg(1999)]{Dejonghe1999} Dejonghe, H., van Caelenberg, K.\ 1999, IAUS, 191, 501
\bibitem[Eggen(1998)]{Eggen1998} Eggen, O. J.,\ 1998, AJ, 115, 2435
\bibitem[Feast et al.(1989)]{Feast1989} Feast, M. ~W., Glass, I. ~S., Whitelock, P. ~A., Catchpole, R. ~M.,\ 1989, MNRAS, 241, 375 
\bibitem[Fleischer et al.(1992)]{Fleischer1992} Fleischer, A. J., Gauger, A., Sedlmayr, E.\ 1992, A\&A, 266, 321
\bibitem[Fluks et al.(1994)]{Fluks1994} Fluks, M. A.; Plez, B.; The, P. S., et al.\ 1994, A\&AS, 105, 311
\bibitem[Fox et al.(1984)]{Fox1984} Fox, M. W., Wood, P. R., Dopita, M. A., 1984, ApJ, 286, 337
\bibitem[Fox \& Wood(1985)]{Fox1985} Fox, M. W., Wood, P. R., 1985, ApJ, 297, 455
\bibitem[Gautschy$-$Loidl et al.(2004)]{Gautschy2004} Gautschy-Loidl, R., H\"ofner, S., J\o rgensen, U. G., Hron, J.\ 2004, A\&A, 422, 289
%\bibitem[Gobrecht et al.(2016)]{Gobrecht2016} Gobrecht, D., Cherchneff, I., Sarangi, A., Plane, J. M. C., Bromley, S. T.\ 2016, A\&A, 585, 6
\bibitem[Groenewegen et al.(2009)]{Groenewegen2009} Groenewegen, M. A. T., Lan\c con, A., Marescaux, M.\ 2009, A \& A, 504, 1031
\bibitem[Groenewegen(2012)]{Groenewegen2012} Groenewegen, M. ~A. ~T.,\ 2012, A\&A, 543, 36
\bibitem[Guha Niyogi et al.(2011)]{Niyogi2011} Guha Niyogi, Suklima, Speck, Angela K., Onaka, T.,\  2011, \apj, 733, 93
\bibitem[Gustafsson et al.(2008)]{Gustafsson2008}  Gustafsson, B., Edvardsson, B., Eriksson, K., et al.\ 2008, A \& A, 486, 951
\bibitem[Habing(1988)]{Habing1988} Habing, H. ~J.\ 1988, A\&A, 200, 40
\bibitem[Habing(1996)]{Habing1996} Habing, H. ~J.\ 1996, A\&AR, 7, 97
\bibitem[Hambsch(2012)]{Hambsch2012} Hambsch, F. ~-J.\  2012, JAAVSO, 40, 1003
\bibitem[Herwig(2005)]{Herwig2005} Herwig, F.\ 2005, ARA\&A, 43, 435
%\bibitem[Hinkle et al.(1976)]{Hinkle1976} Hinkle, K. H., Lambert, D. L., Snell, R. L.,\ 1976, \apj, 210, 684
\bibitem[Hinkle(1978)]{Hinkle1978} Hinkle, K. H.\ 1978, \apj, 220, 210
\bibitem[Hinkle \& Barnes(1979a)]{Hinkle1979a} Hinkle, K. H., Barnes, T. G.\ 1979, \apj, 227, 923
\bibitem[Hinkle \& Barnes(1979b)]{Hinkle1979b} Hinkle, K. H., Barnes, T. G.\ 1979, \apj, 234, 548
\bibitem[Hinkle et al.(1982)]{Hinkle1982} Hinkle, K. H., Hall, D. N. B., Ridgway, S. T.,\ 1982, \apj, 252, 697
\bibitem[Hinkle et al.(1989)]{Hinkle1989} Hinkle, Kenneth H., Wilson, Teresa D., Scharlach, Werner W. ~G., Fekel, Francis C.,\ 1989, AJ, 98, 1820
\bibitem[H\"ofner et al.(1998)]{Hofner1998} H\"ofner, S., J\o rgensen, U. G., Loidl, R., Aringer, B.\ 1998, A\&A, 340, 497
\bibitem[Hughes \& Wood(1990)]{Hughes1990} Hughes, Shaun M. ~G., Wood, Peter R.,\ 1990, AJ, 99, 784
\bibitem[Ireland et al.(2004)]{Ireland2004} Ireland, M. ~J., Scholz, M., Wood, P. ~R.,\ 2004, MNRAS, 352, 318
\bibitem[Ita et al.(2004)]{Ita2004} Ita, Y., Tanab\'e, T., Matsunaga, N., et al.,\ 2004, MNRAS, 353, 705
\bibitem[Ita \& Matsunaga(2011)]{Ita2011} Ita, Y., Matsunaga, N.,\ 2011, MNRAS, 412, 2345
\bibitem[Ishihara et al.(2010)]{Ishihara2010} Ishihara, D., Onaka, T., Kataza, H., et al.,\  2010, A\& A, 514A, 1
\bibitem[Ivezi\'{c} \& Elitzur(1997)]{Ivezic1997} Ivezi\'{c}, \v{Z}, Elitzur, M.,\ 1997, MNRAS, 287, 799
\bibitem[Ivezi\'{c} et al.(1999)]{Ivezic1999} Ivezi\'{c}, \v{Z}, Elitzur, M.,\ 1999, DUSTY user manual, University of Kentucky internal report
\bibitem[Johnson et al.(1968)]{Johnson1968} Johnson, H. L., Coleman, I., Mitchell, R. I., Steinmetz, D. L.,\ 1968, CoLPL, 7, 83
%\bibitem[Joy (1926)]{Joy1926} Joy, A. H.\ 1926, \apj, 63, 281
%\bibitem[Joyce et al.(1998)]{Joyce1998}Joyce, Richard R., Hinkle, Kenneth H., Wallace, Lloyd, Dulick, Michael, Lambert, David L. 1998, AJ, 116, 2520
\bibitem[Jura \& Kleinmann(1990)]{Jura1990} Jura, M., \& Kleinmann, S. G.\ 1990, \apj, 364, 663
\bibitem[Jura \& Kleinmann(1992)]{Jura1992} Jura, M., \& Kleinmann, S. G.\ 1992, ApJS, 79, 105
\bibitem[Juri\'{c} et al.(2008)]{Juric2008}, Juri\'{c}, M., Ivezi\'{c}, \v{Z}., Brooks, A., et al.\ 2008, \apj, 673, 864
%\bibitem[Karovicova et al.(2013)]{Karovicova2013} Karovicova, I., Wittkowski, M., Ohnaka, K., Boboltz, D. A., Fossat, E., Scholz, M.\ 2013, A\&A, 560, 75
\bibitem[Keeley(1970)]{Keeley1970} Keeley, D. A.,\ 1970, \apj, 161, 657
%\bibitem[Kenyon \& Fernandez-Castro (1987)]{Kenyon1987} Kenyon, Scott J., Fernandez-Castro, T.,\ 1987, AJ, 93, 938
\bibitem[Kharchenko et al.(2002)]{Kharchenko2002} Kharchenko, N., Kilpio, E., Malkov, O., Schilbach, E.,\ 2002, A\&A, 384, 925
%\bibitem[Labeyrie et al.(1977)]{Labeyrie1977}  Labeyrie, A., Koechlin, L., Bonneau, D., Blazit, A., Foy, R.,\ 1977, \apj, 218, 75
\bibitem[Lan\c con et al.(1999)]{Lancon1999} Lan\c con, A., Mouhcine, M., Fioc, M., Silva, D.\ 1999, A\&A, 344, 21
\bibitem[Lan\c con \& Wood(2000)]{Lancon2000} Lan\c con, A., \& Wood, P. R.,\ 2000, A\&AS, 146, 217
\bibitem[Lan\c con et al.(2007)]{Lancon2007} Lan\c con, A., Hauschildt, P. H., Ladjal, D., Mouhcine, M.,\ 2007, A\&A, 468, 205
%\bibitem[ Lattanzio \& Wood (2004)]{Lattanzio2004} Lattanzio, J. ~C., \& Wood, P.\ 2004, in Asymptotic Giant Branch Stars, ed. H. J. Habing, \& H. Olofsson (Springer), chap. 2, 23
\bibitem[Lebzelter et al.(1999)]{Lebzelter1999} Lebzelter, Th., Hinkle, Kenneth H., Hron, J.,\ 1999, A\&A, 341, 224
%\bibitem[Lebzelter et al.(2006)]{Lebzelter2006} Lebzelter, Th., Posch, Th., Hinkle, K., Wood, P. R., \& Bouwman, J.,\ 2006, \apj, 653L, 145
%\bibitem[Lebzelter et al. (2010)]{Lebzelter2010} Lebzelter, T., Nowotny, W., H\"ofner, S., Lederer, M. T., Hinkle, K. H., Aringer, B.,\ 2010, A\&A, 517, 6
\bibitem[Le Bertre(1992)]{LeBertre1992}  Le Bertre, T.,\ 1992, A\& AS, 94, 377
\bibitem[Lenz \& Breger(2005)]{Lenz2005} Lenz, P., Breger, M.\ 2005, CoAst, 146, 53
%\bibitem[Lipunov et al.(2010)]{Lipunov2010} Lipunov, V., Kornilov, V., Gorbovskoy, E., et al.\ 2010, AdAst, 2010, 30	
\bibitem[Lipunov et al.(2016)]{Lipunov2016} Lipunov, V., Gorbovskoy, E., Afanasiev, V., et al.\ 2016, A\&A, 588, 90	
\bibitem[Lipunov et al.(2017)]{Lipunov2017} Lipunov, V. M., Kornilov, V., Gorbovskoy, E., et al.\ 2017, MNRAS, 465, 3656
\bibitem[Lockwood (1972)]{Lockwood1972} Lockwood, G. W.,\ 1972, \apjs, 24, 375
\bibitem[Loidl et al.(1999)]{Loidl1999} Loidl, R., H\"ofner, S., J\o rgensen, U. G., Aringer, B.\ 1999, A\&A, 342, 531
\bibitem[Lomb (1976)]{Lomb1976} Lomb, N. R.,\ 1976, Ap\&SS, 39, 447
%\bibitem[Mathis et al.(1977}]{Mathis1977}  Mathis, J. S., Rumpl, W., Nordsieck, K. H., 1977, \apj, 217, 425
\bibitem[Mattei(1997)]{Mattei1997} Mattei, J. A.,\ 1997, JAAVSO, 25, 57
\bibitem[Merrill(1962)]{Merrill1962} Merrill, Paul W., Deutsch, Armin J., Keenan, Philip C.\ 1962, \apj, 136, 21
\bibitem[Mondal et al.(2005)]{Mondal2005} Mondal, Soumen, Chandrasekhar, T.,\ 2005, AJ, 130, 842
\bibitem[Mondal et al.(2010)]{Mondal2010} Mondal, S., Lin, C. C., Chen, W. P., et al,\ 2010, AJ, 139, 2026
\bibitem[Mondal et al.(2013)]{Mondal2013} Mondal, S., Das, R. K., Ashok, N. ~M., Banerjee, D. ~P. ~K., Dutta, S., Ghosh, S., Mondal, A.,\ 2013, ATel, \#4931
\bibitem[Ninan et al.(2014)]{Ninan2014} Ninan, J. P., Ojha, D. K., Ghosh, S. K., et al.,\ 2014, Journal of Astronomical Instrumentation, 3, 1450006
\bibitem[Ngeow et al.(2013)]{Ngeow2013}  Ngeow Chow-Choong, Lucchini S. ,  Kanbur S.,  Barrett B., Lin B.,\ 2013, arXiv:1309.4297 [astro-ph.SR]
\bibitem[Nowotny et al.(2005a)]{Nowotny2005a} Nowotny, W., Aringer, B., H\"ofner, S., Gautschy-Loidl, R., Windsteig, W. 2005, A\&A, 437, 273
\bibitem[Nowotny et al.(2005b)]{Nowotny2005b} Nowotny, W., Lebzelter, T., Hron, J., H\"ofner, S. 2005, A\&A, 437, 285
\bibitem[Nowotny et al.(2010)]{Nowotny2010} Nowotny, W., H\"ofner, S., Aringer, B.\ 2010, A\&A, 514A, 35N
\bibitem[O'Connell(1973)]{Connell1973} O'Connell, R. ~W.,\ 1973, AJ, 78, 1074
\bibitem[Olofsson(2004)]{Olofsson2004} Olofsson, H.\ 2004, in Asymptotic Giant Branch Stars, ed. H. J. Habing, \& H. Olofsson (Springer), Chap. 7, 325
%\bibitem[Origlia et al. (1993)]{Origlia1993} Origlia, L., Moorwood, A. ~F. ~M., Oliva, E.,\ 1993, A\&A, 280, 536O
%\bibitem[Ossenkopf et al.(1992}]{Ossenkopf1992} Ossenkopf, V., Henning, Th., Mathis, J. S., 1992, A\&A, 261, 567O
\bibitem[Perrin(2004)]{Perrin2004} Perrin, G., Ridgway, S. ~T., Mennesson, B., et al.,\ 2004, A\&A, 426, 279
\bibitem[Ramirez et al.(1997)]{Ramirez1997} Ramirez, S. ~V., Depoy, D. ~L., Frogel, Jay A., Sellgren, K., Blum, R. ~D.,\ 1997, AJ, 113, 1411
\bibitem[Rayner et al.(2009)]{Rayner2009} Rayner, John T., Cushing, Michael C., Vacca, William D.,\ 2009, \apjs, 185, 289
\bibitem[Reid \& Goldston(2002)]{Reid2002}  Reid M. ~J., Goldston  J. ~E.,\ 2002, \apj, 568, 931 
%\bibitem[Rejkuba et al.(2007)]{Rejkuba2007} Rejkuba, M., Dubath, P., Minniti, D., Meylan, G.,\ 2007, A\&A, 469, 147
\bibitem[Rojas-Ayala et al.(2012)]{Rojas2012} Rojas-Ayala, Bárbara, Covey, Kevin R., Muirhead, Philip S., Lloyd, James P.,\ 2012, \apj, 748, 93R
%\bibitem[Sacuto et al.(2013)]{Sacuto2013} Sacuto, S., Ramstedt, S., H\"ofner, S., Olofsson, H., Bladh, S., Eriksson, K., Aringer, B., Klotz, D., Maercker, M.\ 2013, A\&A, 551, 72
\bibitem[Scargle(1982)]{Scargle1982} Scargle, J. ~D.,\ 1982, \apj, 263, 835S
\bibitem[Schlafly et al.(2011)]{Schlafly2011} Schlafly, Edward F., Finkbeiner, Douglas P.,\ 2011, \apj, 737, 103
\bibitem[Schlegel et al.(1998)]{Schlegel1998}  Schlegel, David J., Finkbeiner, D. P., Davis, M.,\ 1998, \apj, 500, 525	
\bibitem[Scholz et al.(2012)]{Scholz2012} Scholz, A., Muzic, K., Geers, V., et al.,\ 2012, \apj, 744, 6
%\bibitem[Schwarz(1978)]{Schwarz1978} Schwarz, G.\ 1978, Ann. Stat., 6, 461 
%\bibitem[Sharpless (1956)]{Sharpless1956} Sharpless, S.\ 1956, \apj, 124, 342
\bibitem[Smith et al. (2002)]{Smith2002} Smith, B. ~J., Leisawitz, D., Castelaz, M. ~W., Luttermoser, Donald,\ 2002, \aj, 123, 948
\bibitem[Smith et al.(2006)]{Smith2006} Smith, B. ~J., Price, S. ~D., Moffett, A. ~J.,\ 2006, AJ, 131, 612
\bibitem[Soszy\'nski et al.(2012)]{Soszynski2012} Soszy\'nski, I., Udalski, A., Poleski, R., Koz\l owski, S., Wyrzykowski, \L., Pietrukowicz, P., Szyma\'nski, M. K., Kubiak, M., Pietrzy\'nski, G., Ulaczyk, K., Skowron, J.,\ 2012, AcA, 62, 219
\bibitem[Strecker et al.(1978)]{Strecker1978} Strecker, D. ~W., Erickson, E. ~F., Witteborn, F. ~C.,\ 1978, \aj, 83, 26
\bibitem[Tej et. al.(2003a)]{Tej2003a}Tej, A., Lan\c on, A., Scholz, M. \&  Wood, P. ~R.\ 2003a, A\&A, 412, 481
\bibitem[Tej et. al.(2003b)]{Tej2003b}Tej, A., Lan\c on, A. \&  Scholz, M.\ 2003b, A\&A, 401, 347
\bibitem[Tiurina et al.(2013)]{Tiurina2013} Tiurina, N. et al., 2013, \#ATel, 4888
\bibitem[Thompson et al. (2002)]{Thompson2002} Thompson, R. ~R., Creech-Eakman, M. ~J., van Belle, G. ~T.\ 2002, \apj, 577, 447T
%\bibitem[Tsuji(2000)]{Tsuji2000} Tsuji, T.,\ 2000, \apj, 540, 99 
\bibitem[Tsuji (2009)]{Tsuji2009} Tsuji, T.\ 2009, A\&A, 504, 543
\bibitem[Wallace \& Hinkle (1996)]{Wallace1996} Wallace, L., Hinkle, K.,\ 1996, \apjs, 107, 312
\bibitem[Weights et al.(2009)]{Weights2009} Weights, D. ~J., Lucas, P. ~W., Roche, P. ~F., Pinfield, D. ~J., Riddick, F.,\ 2009, MNRAS, 392, 817
\bibitem[Whitelock, Marang \& Feast(2000)]{Whitelock2000} Whitelock, P., Marang, F., Feast, M.,\ 2000, MNRAS, 319, 728
\bibitem[Whitelock et al.(2003)]{Whitelock2003} Whitelock, Patricia A., Feast, M. ~W., van Loon, Jacco Th., Zijlstra, Albert A.,\ 2003, MNRAS, 342, 86W
\bibitem[Whitelock et al.(2008)]{Whitelock2008} Whitelock, Patricia A., Feast, Michael W., van Leeuwen, Floor,\ 2008, MNRAS, 386, 313W
\bibitem[Winters et al. (1997)]{Winters1997} Winters, J. M., Fleischer, A. J., Le Bertre, T., Sedlmayr, E.,\ 1997, A\&A, 326, 305
\bibitem[Winters et al.(2000)]{Winters2000} Winters, J. M., Le Bertre, T., Jeong, K. S., Helling, Ch., Sedlmayr, E.\ 2000, A\&A 361, 641
\bibitem[Wittkowski et al.(2007)]{Wittkowski2007} Wittkowski, M., Boboltz, D. A., Ohnaka, K., Driebe, T., Scholz, M.\ 2007, A\&A,470, 191
\bibitem[Wittkowski et al.(2008)]{Wittkowski2008} Wittkowski, M., Boboltz, D. A., Driebe, T., Le Bouquin, J.-B., Millour, F., Ohnaka, K., Scholz, M.\ 2008, A\&A, 479, 21
\bibitem[Wood \& Sebo(1996)]{Wood1996} Wood, P. R., Sebo, K. M.,\ 1996, MNRAS, 282, 958
\bibitem[Wood et al.(1999)]{Wood1999} Wood, P. R., Alcock, C., Allsman, R. A., et al.\ 1999, IAUS, 191, 151
\bibitem[Wright et al.(2009)]{Wright2009} Wright, N. ~J., Barlow, M. ~J., Greimel, R., Drew, J. ~E., Matsuura, M., Unruh, Y. ~C., Zijlstra, A. ~A.,\ 2009, MNRAS, 400, 1413
%\bibitem[Wright et al.(2010}]{Wright2010} Wright, Edward L. et al., 2010, AJ, 140, 1868
%\bibitem[Zhou (1991)]{Zhou1991} Zhou, Xu,\ 1991, A\&A, 248, 367
\bibitem[Zhu et al.(1999)]{Zhu1999} Zhu, Z. ~X., Friedjung, M., Zhao, G., Hang, H. ~R., Huang, C. ~C.,\ 1999, A\& AS, 140, 69

\end{thebibliography}
\end{document}